\documentclass[useAMS]{mn2e}
\usepackage{graphicx}
\usepackage{epsfig}

\def\ltsima{$\; \buildrel < \over \sim \;$}
\def\lsim{\lower.5ex\hbox{\ltsima}}
\def\gtsima{$\; \buildrel > \over \sim \;$}
\def\gsim{\lower.5ex\hbox{\gtsima}}
\def\mdot {\dot M}
\newcommand{\be}{\begin{equation}}
\newcommand{\en}{\end{equation}}
\newcommand{\ergs}{\rm \ erg \; s^{-1}}

\def\msole {~M_{\odot}}

\begin{document}
\title[]{Mining the Aql X-1 long term X--ray light curve}

\author[Campana, Coti Zelati \& D'Avanzo]{S. Campana$^{1,}$\thanks{E-mail: sergio.campana@brera.inaf.it}, F. Coti Zelati$^{1,2}$, 
P. D'Avanzo${^1}$ \\
$^1$ INAF-Osservatorio astronomico di Brera, Via Bianchi 46, I--23807, Merate (LC), Italy\\
$^2$ Universit\`a di Milano-Bicocca, Piazza della Scienza 3, I--20126 Milano, Italy
}

\maketitle

\begin{abstract}
Aql X-1 is the prototypical low mass X--ray binary transient.  The Rossi X--ray Timing Explorer All Sky
Monitor provided a $\sim 16$ yr coverage revealing 20 outbursts. This is by far the most extensive legacy 
of outbursts from the same source. We investigated the outbursts characteristics in terms of energetics, 
peak luminosities, durations, decays and recurrence times. 
We found  that bright outbursts (peak luminosity $\gsim 10^{37}$ erg s$^{-1}$) equal in number dimmer outbursts 
($\lsim 10^{36.6}$ erg s$^{-1}$). The peak luminosity does not correlate with outburst energetics, durations or 
quiescent times. 
We analysed the latest stages of the outbursts searching for exponential and/or linear decays. Light curve modeling 
led to constraints on the outer disk radius and enabled us to estimate the viscosity and the irradiation parameters. 
The former is larger than what has been obtained for other, shorter orbital period, transients, while the latter  
is somewhat smaller. 
This might be related to the longer orbital period of Aql X-1 with respect to other transient X--ray binaries.
\end{abstract}


\section{Introduction}

Aql X-1 is the most prolific neutron star low mass X--ray binary transient (LMXTs). Seven outbursts were observed by the {\it Vela 5B} satellite 
between 1969--1976 (Priedhorsky \& Terrell 1984). The {\it Ariel 5}  satellite detected three outbursts in the 1976--1979 
time period. Three more outbursts were detected by the All-Sky Monitor (ASM) on board the {\it Ginga} satellite in the 1987--1992
time frame (Kitamoto et al. 1993). Despite this high number of events, the sky coverage was not complete and some 
outbursts might have been lost. Interestingly, the recurrence period of the outbursts was short in the seventies, around 125 d 
(Priedhorsky \& Terrell 1984), but it became longer in the eighties, around 300 d (Kitamoto et al. 1993).

The Rossi X--ray Timing Explorer ($RXTE$, 1996--2012, Levine et al. 1996) and more recently the Monitor All-sky X--ray 
Image ($MAXI$, 2009--now, Matsuoka et al. 2009) All-sky Monitors are continuously surveying with higher sensitivity the X--ray sky. 
Up to Sept. 2011, the $RXTE$/ASM detected twenty outbursts from Aql X-1 (two in common with $MAXI$).
These data provide a unique database for investigations on outburst recurrence times and energetics.

The limiting sensitivity of $RXTE$/ASM is such that in the case of Aql X-1 we can follow the outburst decay down to a 2--10 keV 
luminosity of $\sim 5\times 10^{35}\ergs$ (at a distance of 5 kpc; Jonker \& Nelemans 2004).
In the case of $MAXI$ we can reach a somewhat better (within a factor of two) limiting sensitivity.
In addition to the possibility of detecting faint outbursts, this luminosity is sufficiently faint to allow us to investigate the outburst's decay law.
King \& Ritter (1998) showed that X--ray heating during the decay from an outburst causes the light curves of LMXTs
to show either exponential or linear decays, depending on whether or not the X--ray luminosity 
is sufficiently high to keep the outer disc edge hot. The rationale behind this is that if the disc's outer portion is hot the entire 
disc is hot and matter can be efficiently transferred to the compact object, delivering a fast drain of the accretion disc.

In this paper we exploit the $RXTE$/ASM (and $MAXI$) database on Aql X-1. In Section 2 we discuss the data and characterise the
outbursts. In Section 3 we focus on correlations among outburst properties and we fit the decay profiles of the outbursts to derive the outer 
disc radius (Section 4). In Section 5 we discuss our results.

\section{Data analysis}

\subsection{$RXTE$ data}

\begin{figure}
\begin{center}
\includegraphics[width=6cm,angle=-90]{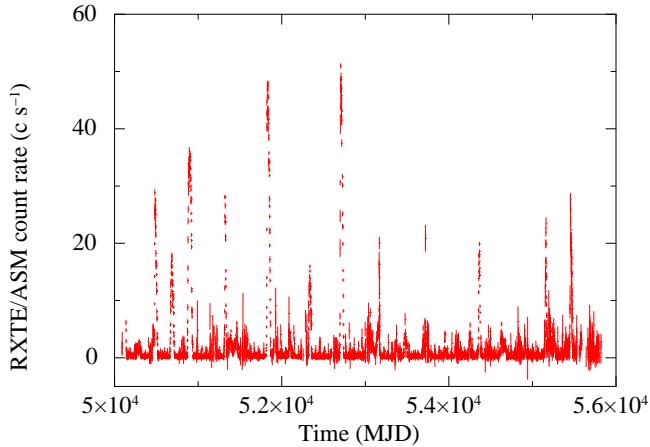}
\end{center}
\caption{$RXTE$/ASM light curve of Aql X-1. Data are one day averages and report the ASM rate in the 1.5--12 keV energy band.
Data have been filtered in order to get rid of negative values ($\sim 28\%$ of the data). 
For comparison the Crab nebula provides 75 count s$^{-1}$.}
\label{rxte}
\vskip -0.1truecm
\end{figure}

The $RXTE$/ASM consisted of three position-sensitive Scanning Shadow Cameras (Levine et al. 1996). $RXTE$ started operating 
in Jan. 1996. The satellite was slowly spinning, observing sources for 90 s (dwell) during the 96 min duration of each orbit. 
Raw data points provided the fitted source rates in the 1.5--12 keV energy band from one dwell.
One-day averages are more commonly used, resulting from the average of the fitted source rates from the dwells during that time interval 
(typically 5-10). Each measurement (count rate) has an error associated, coming from the quadrature average of the estimated 
errors on the individual dwells. The one-day average light curve of Aql X-1 is shown in Fig. \ref{rxte} (the flux of the Crab is 
$\sim 75$ count s$^{-1}$).

\begin{figure}
\begin{center}
\includegraphics[width=6cm,angle=-90]{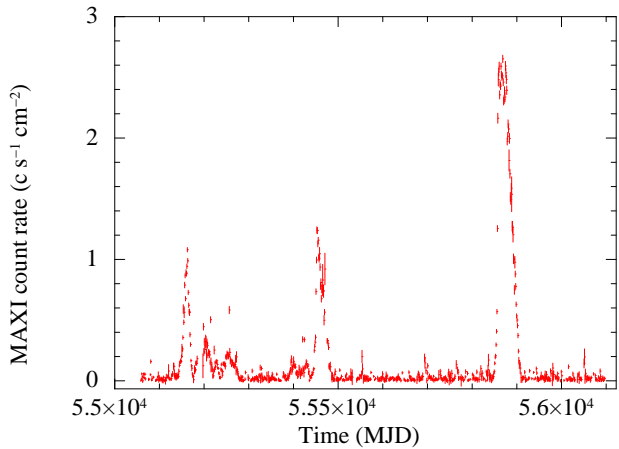}
\end{center}
\caption{$MAXI$ light curve of Aql X-1. Data are one day averages in the 2--20 keV energy band.
Data have been filtered in order to get rid of negative values ($\sim 20\%$ of the data). 
For comparison the Crab nebula provides 3.5 count s$^{-1}$ cm$^{-2}$.}
\label{maxi}
\vskip -0.1truecm
\end{figure}

\subsection{$MAXI$ data}

The $MAXI$ is a position-sensitive X--ray all-sky monitor instrument, to scan almost the entire sky 
during its orbit, once every 96 minutes (Matsuoka et al. 2009). $MAXI$ is located on the International Space Station and is collecting 
data since Aug. 2009.  The detection sensitivity is somewhat better than $RXTE$/ASM, reaching a few mCrab during one day observations. 
The operational energy range is 2--20 keV. The one-day average light curve of Aql X-1 is shown in Fig. \ref{maxi} (the flux of the 
Crab is $\sim 3.5$ count s$^{-1}$ cm$^{-2}$).

\begin{figure}
\begin{center}
\includegraphics[width=6cm,angle=-90]{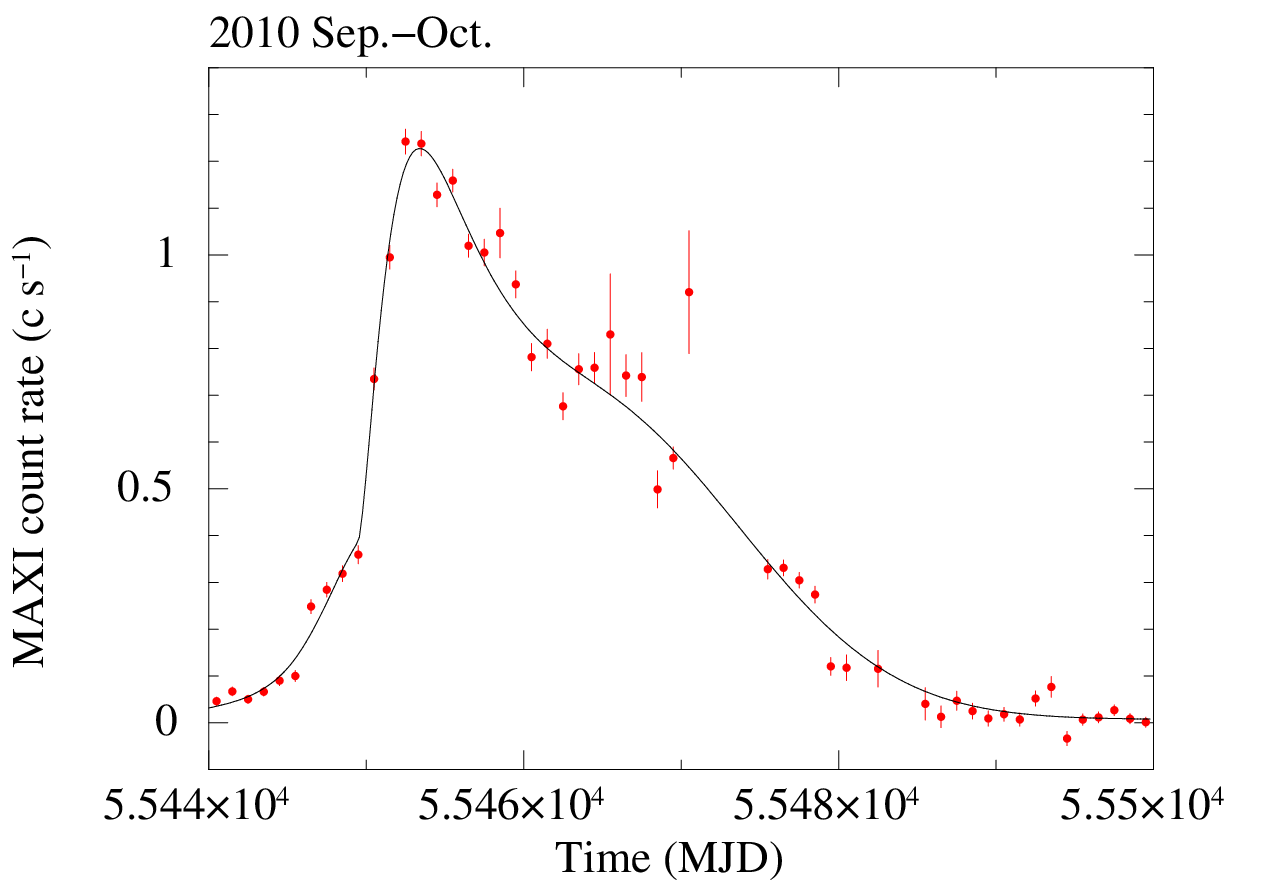}
\includegraphics[width=6cm,angle=-90]{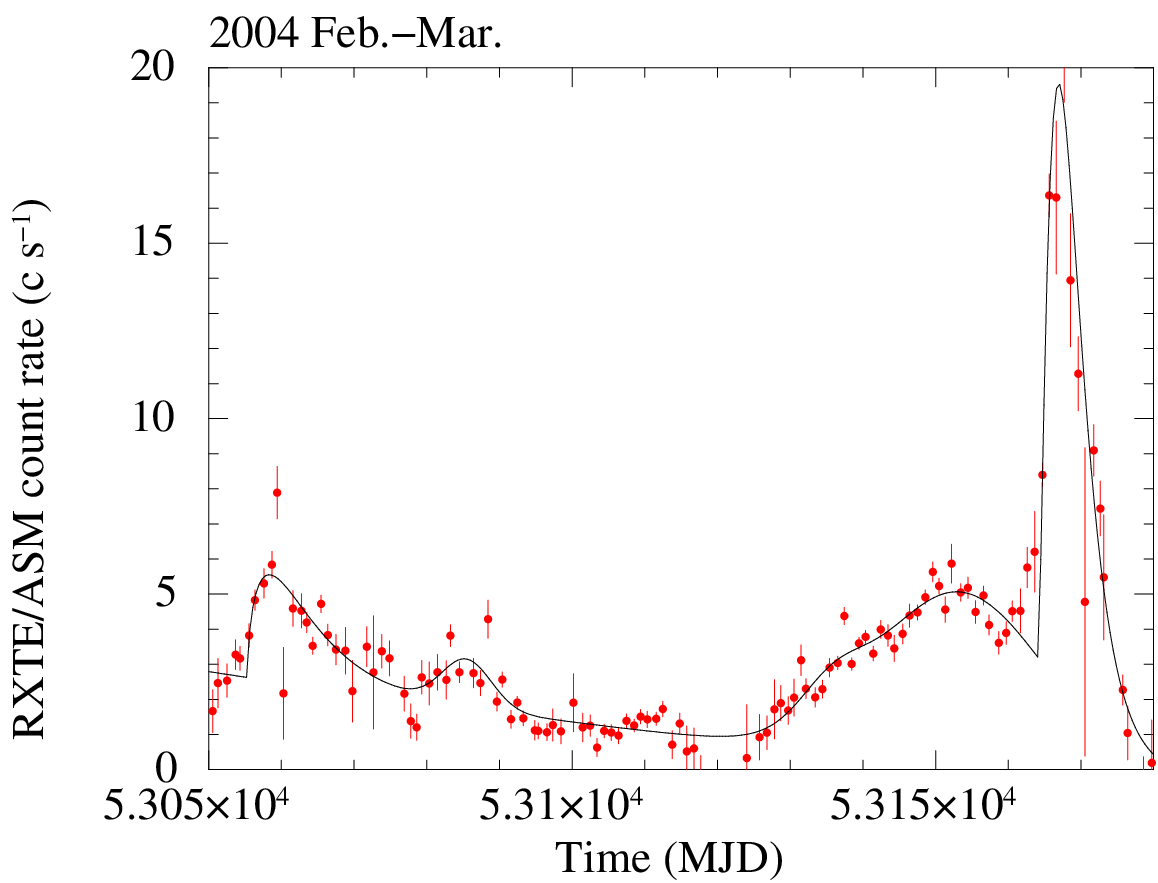}
\includegraphics[width=5.5cm,angle=-90]{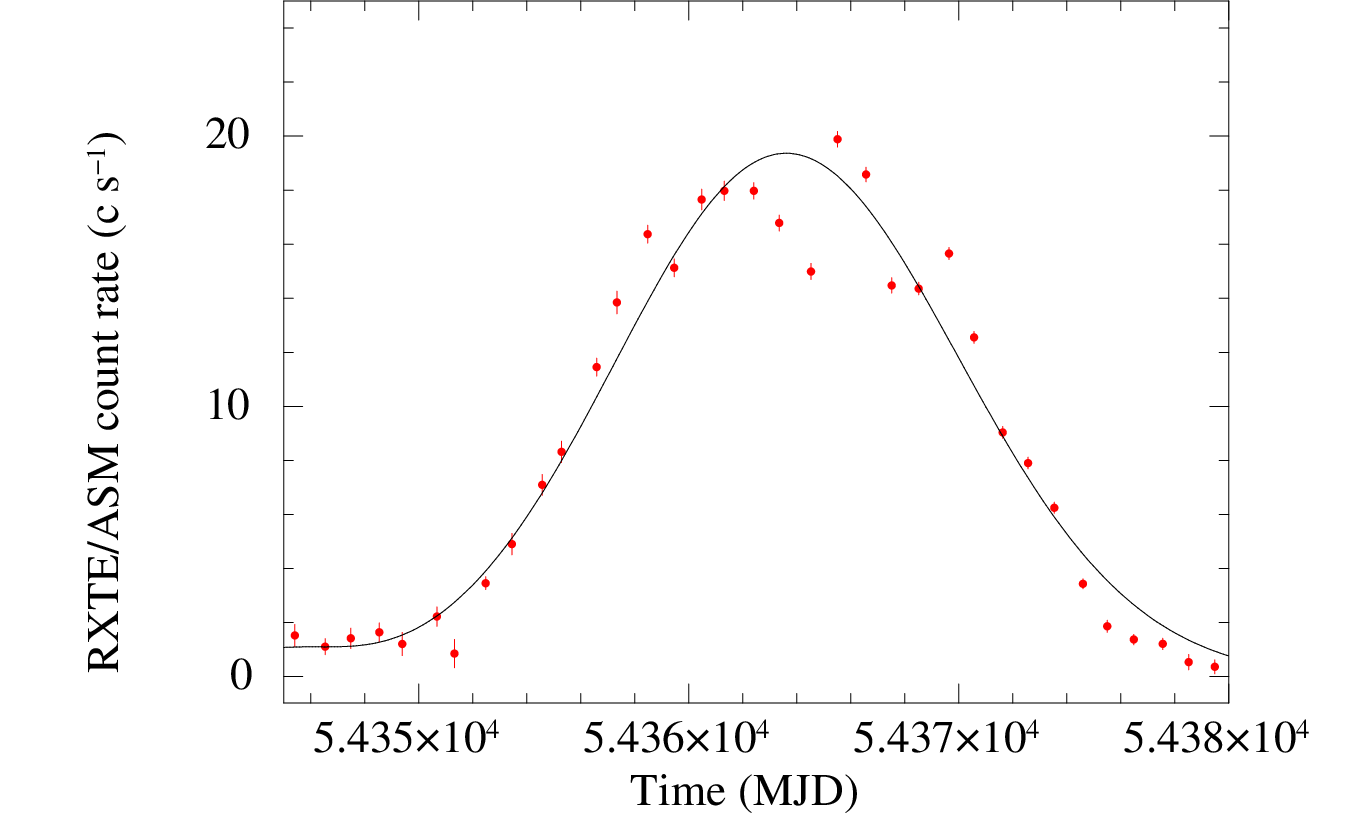}
\end{center}
\caption{Examples of outburst fitting. From top to bottom are shown the Aql X-1 outburst of 2010 Sep-Oct ($MAXI$, FRED),
2004 Feb-Mar ($RXTE$, LIS-FRED) and 2007 Sep ($RXTE$, Gaus). }
\label{outb}
\vskip -0.1truecm
\end{figure}

\begin{table*}
\caption{Aql X-1 outbursts.}
\begin{center}
\begin{tabular}{ccccccccc}
\hline
Outburst         & Morphology  & Duration & Day Peak & Rate peak & Rise & Decay & Quiescent time$^*$ & Energy \\ 
                         &                        & (d)            & (MJD)        & (mCrab)   & (d)     & (d)       &   (d)                      & ($10^{42}$ erg)\\
\hline
1996 Feb         & FRED           & 9              & 50136       & 87               & 1       &   8       & --     &  1.1 \\
1996 May-Aug & Multi-peak  & 95           & 50298       & 38               & 15     & 21       & 81   &  7.6 \\
1997 Jan-Mar & LIS-FRED-P & 90          & 50483       & 379             & 8      & 30        & 120 &  21.7 \\
1997 Aug-Sep & Gaus-P        & 64           & 50684       & 220             & 28    & 36        & 126 & 12.6 \\
1998 Feb-May & Multi-peak   & 85           & 50885      & 482              & 19   & 40        & 145  & 27.6\\
1999 May-Nov & FRED-P-LIS& 184        & 51320       & 376             & 12    & 172     & 358  & 21.5 \\
2000 Sep-Nov & FRED-P      & 58            & 51841      & 628              & 12   & 28        & 318  & 36.0\\
2001 Jun-Aug & Multi-peak  & 58            & 52087       & 65                & 16   & 42        & 203  & 3.7\\
2002 Feb-Apr & FRED-P       & 60            & 52337       & 209              & 27   & 33        & 181  & 12.0\\
2003 Feb-Apr & FRED           & 64            & 52706       & 649             & 20   & 44         & 316  & 37.1\\
2004 Feb-Jun & LIS-FRED   & 130          & 53167       & 259              & 8     & 13        & 300  & 14.8\\
2005 Mar-May & FRED-P      & 70            & 53473       & 66                & 23   & 47        & 270  & 3.8\\
2005 Nov         & FRED           & 24           & 53692       & 68                & 6     & 12         & 160  & 3.9\\
2006 Jul-Aug  & FRED-P       & 40           & 53952       & 57                & 7     & 16         & 236  & 3.3\\
2007 May-Jun & Multi-peak  & 54           & 54254       & 64                 & 24  & 30         & 250  & 3.7\\
2007 Sep         & Gaus            & 36           & 54364       & 257              & 24  & 12         &  56    & 14.7\\
2008 May-Jun & FRED-P      & 150         & 54627       & 55                & 32   & 73        & 174   & 3.1\\
2009 Mar-Apr & FRED           & 40           & 54906       & 44                & 6     & 34         & 200   & 2.5\\
2009 Nov-Dec & Gaus-P       & 38           & 55163      & 311              & 23  & 15          & 200   & 24.9\\
2010 Sep-Oct & FRED           & 60           & 55453      & 349              & 13   & 47         & 262   & 27.8\\
\hline
\end{tabular} 
\end{center}
{\leftline{$^*$ Quiescent time before the start of the outburst.}}
\end{table*}

\section{Outburst analysis}

During the Jan 1996 -- Sep 2011 time interval 20 outbursts from Aql X-1 were recorded by $RXTE$ and $MAXI$.
Usually the morphology of LMXTs is characterised by a fast rise and an exponential decay shape (Chen et al. 1997).
In systems like Aql X-1, i.e. neutron star X--ray binary transients, the thermal viscous disc instability model (DIM; Lasota 2001) is 
usually considered  to explain the transient behaviour. However, it has been shown that the DIM alone cannot reproduce FRED-like 
outbursts. The other key ingredient is X--ray irradiation. The irradiation of the outer parts of the accretion disc is caused by the intense 
X--ray flux produced from the innermost disc regions and from the neutron star itself. Irradiation dominates and heavily changes the 
outer disc temperature profile. 

X--ray irradiation is also responsible for the decay of the outburst. If the outer disc edge is kept hot (i.e. above the temperature
of the DIM hot viscous state), accretion disc theory predicts an exponential decay for the outburst (King \& Ritter 1998). 
On the contrary if (or when) disc irradiation is not strong enough, the outer disc edge temperature falls below the limiting 
hot temperature of the DIM, and we can expect a linear decay of the X--ray flux.

We have analysed and characterised all the 20 Aql X-1 outbursts observed by $RXTE$/ASM.
For each one of them we have first classified the outbursts, following Maitra \& Baylin (2008).
Outbursts were classified based on their X--ray morphology as: $i$) Fast Rise Exponential Decay (FRED); $ii$) FRED with peak(s):
when overlaid to the FRED morphology  there are pronounced peaks; $iii$) multi-peaks: when the presence of many peaks 
distorts the overall morphology; $iv$) Gaussian (Gaus): when the rise and decay times are more similar. 
The typical observed ($e$-folding) FRED rise times are approximately few days (up to a week) and the decay times are 3--4 weeks.
According to Wachter et al. (2002), we  observe that usually, following an outburst (but sometimes also preceding an outburst), the source
might remain stuck in a low-intensity state (LIS), before ending (starting) the outburst. 
Outbursts are classified according to this scheme in Table 1. A few examples are shown in Fig. \ref {outb}.

We then investigated outburst characteristic times. The duration of an outburst is simply computed as the time difference between 
the first and last detection of a given outburst detected by the $RXTE$/ASM or $MAXI$ (the latest two, see Table 1).
The day of the outburst peak is reported in Table 1. We also consider rise and decay times.  In order to be model independent, we prefer 
to separately indicate the time needed to reach the outburst peak (rise) and the time needed to reach non-detection (decay). 
In the case of a typical FRED-like 
(or Gaussian-like) outburst, the sum of the rise and decay times gives the outburst duration time. This is not the case when other 
main peaks are present or when there is a LIS.

We fitted each outburst profile with simple models in order to get a description of their overall shapes. A few of them are shown in Fig. \ref{outb}.
It is clear that outbursts can be very complicated and the analytical description does not provide a detailed account of all the small 
variations observed in the light curve. For this reason the reduced $\chi^2$ obtained from the fit are often much larger than unity.
Despite this, we verified that the overall fit we considered is able to account for the total energetics of the outbursts to a $\sim 10\%$  precision
(this has been done for a few outburst cases, integrating point by point the light curve and extrapolating the data when observations were missing
and comparing the results with the integral of the analytical function).
To derive the outburst energetics we simply assumed a Crab-like spectrum.
Knowing the source distance (5 kpc) and the column density ($N_H\sim (3-4)\times10^{21}$ cm$^{-2}$, e.g. 
Sakurai et al. 2012), one can convert count rates into luminosities.

\section{Outer disc radius estimate}

The late-stage light curve of LMXT outbursts are often characterised in terms of exponential or linear decays.
If the outer disc edge is kept hot (i.e. above the temperature of the DIM hot viscous state), accretion disc theory predicts 
an exponential decay for the outburst, otherwise the decay is predicted to be linear (King \& Ritter 1998). King \& Ritter (1998) 
noticed that the exponential decay must revert to the linear mode when the X--ray flux has decreased sufficiently, but did not analyse 
this in detail.  
Powell, Haswell \& Falanga (2007) took into account the continuous mass transfer into the disk from the donor star, and rewrote the 
total central mass accretion rate equation for the decay outburst profiles. 
When the maximum disk radius of the hot viscous portion of the disk, $R_h$, becomes smaller than the total disk radius, $R_{\rm disk}$, 
(i.e. when the outer part of the disk is no longer kept in the hot, viscous state by central irradiation), then $R_h$ will decrease linearly 
and, correspondingly the X--ray luminosity will turn from an exponential to a linear decay.
When this occurs a distinctive knee in the light curve outburst decay can be expected (see also Shahbaz, Charles \& King 1998). 

\begin{figure}
\begin{center}
\includegraphics[width=6cm,angle=-90]{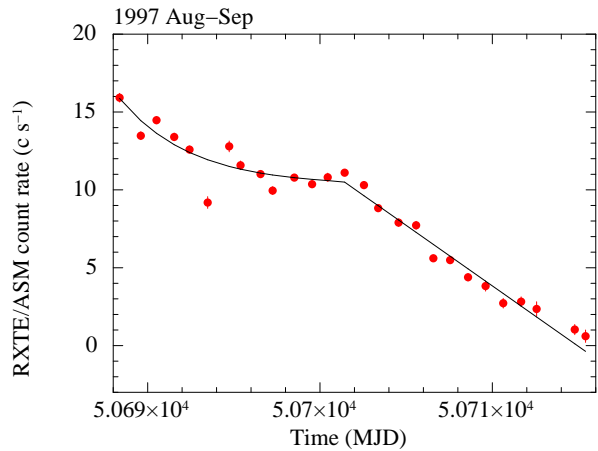}
\end{center}
\caption{Example of a fit with an exponential and linear function of the 1997 Aug-Sep outburst from Aql X-1 ($RXTE$/ASM data).
The two curves join at the exponential knee.
 }
\label{explin}
\vskip -0.1truecm
\end{figure}

At variance with the above outburst light curve fits, here we fit all the latest stages of Aql X-1 outbursts with a function 
\begin{equation}
  N=(N_t-N_e)\exp\left(-\frac{t-t_t}{\tau_e}\right)+N_e
\end{equation}
for the exponential part, where $N_e$ is the count rate corresponding to the limit of the 
exponential decay, $N_t$ is the count rate at the knee, and $\tau_e$ is the
time-scale of the exponential decay.  The linear decline is given by:
\begin{equation}
  N=N_t\left(1-\frac{t-t_t}{\tau_l}\right),
\end{equation}
where $\tau_l$ is the time after the knee at which the (extrapolated) count rate
goes to zero. The two functions join at the exponential/linear transition knee.
This function fully exploits Powell et al. (2007) outburst decay theory.
Based on a Crab-like spectrum, one can convert rates into luminosities and
relate the outburst decay to physical quantities of the system.  An example is shown in Fig. \ref{explin}.

The rate $N_t$ can be converted into a luminosity 
$L_t$ and related to the outer disc radius as $L_t=R_{\rm disc}(L_t)^2/\Phi$, where $\Phi$ is a constant related to the disc 
irradiation properties. Current estimates of $\Phi$ are in the range  $(1-9)\times 10^{-15}$ cm$^2$ s erg$^{-1}$
(see Eq. 2 in Powell et al. 2007 and King \& Ritter 1998). 
A different estimate of the outer disc radius comes from the exponential decay time.
This is related to the radius as $\tau_e={{R_{\rm disc}(\tau_e)^2}\over {3\,\nu}}$, where $\nu$ is 
the viscosity at the outer disc edge. An estimate of the viscosity parameter $\nu$ is in the range $(4-10)\times 10^{14}$ cm$^2$ s$^{-1}$ 
(Powell et al. 2007; King \& Ritter 1998).

Additional constraints on the outer disc boundary come from the fact that it must be larger than the circularisation radius 
(the radius at which matter in Keplerian orbit has the same angular momentum as at the first Lagrangian point) 
and smaller than the first Lagrangian point from the center of the primary. 
In the case of Aql X-1 with an orbital period of 18.97 hr (Welsh, Robinson \&Young 2000), 
the circularisation radius is $R_{\rm circ}\sim 5.0\times 10^{10}$ cm
(assuming a mass ratio of 0.3) and the distance of the first Lagrangian point from the center 
of the primary is $b_1=2.0\times 10^{11}$ cm (Frank, King \& Raine 2002). 

\begin{table*}
\caption{Aql X-1 outburst decay fits.}
\begin{center}
\begin{tabular}{ccccccc}
\hline
Outburst         & Time interval      & $N_t$                        & $N_e$                        & $\tau_e$                   &$R_{\rm disc}(L_t)$  &$R_{\rm disc}(\tau_e)$\\
                         &      (MJD)             &(c s$^{-1}$)               &(c s$^{-1}$)                 & (d)                              &  ($10^{10}$ cm)       & ($10^{10}$ cm) \\
\hline
1996 Feb          & too few points&	--                             &      --                             &       --                          &      --                            &     -- \\
1996 May-Aug & too few points &	--                             &      --                             &       --                          &      --                            &     -- \\
1997 Jan-Mar  & 50503--50522&$15.4^{+0.3}_{-0.2}$&$14.5^{+0.1}_{-3.5}$&$4.5^{+4.6}_{-2.8}$ &$14.0^{+0.1}_{-0.1}$&$10.1^{+4.3}_{-3.9}$\\	
1997 Aug-Sep & 50708--50719&$10.5^{+0.3}_{-0.3}$&$10.6^{+0.7}_{-0.9}$&$4.2^{+1.2}_{-0.8}$ &$11.5^{+0.2}_{-0.2}$&$9.8^{+1.3}_{-1.0}$\\	
1998 Feb-May & 50922--50949&$29.2^{+0.2}_{-0.2}$&$29.2^{+0.2}_{-0.2}$&$1.1^{+0.1}_{-0.1}$ &$19.2^{+0.1}_{-0.1}$&$5.0^{+0.1}_{-0.1}$\\
1999 May-Nov & 51330--51347&$13.1^{+0.3}_{-0.3}$&$0.0^{+2.5}_{-0.0}$&$8.2^{+0.2}_{-0.7}$   &$12.9^{+0.1}_{-0.2}$&$13.7^{+0.2}_{-0.6}$\\
2000 Sep-Nov & 51857--51875&$32.1^{+1.1}_{-1.0}$&$19.9^{+0.0}_{-0.1}$&$15.3^{+0.1}_{-5.6}$&$20.1^{+0.3}_{-0.3}$&$18.7^{+0.1}_{-3.8}$\\
2001 Jun-Aug & too few points  &	--                             &      --                             &       --                          &      --                            &     -- \\
2002 Feb-Apr & 52354--52367 &$8.3^{+0.4}_{-0.4}$&$8.0^{+0.5}_{-0.8}$    &$4.4^{+1.2}_{-0.9}$    &$10.3^{+0.2}_{-0.3}$&$10.0^{+1.3}_{-1.1}$\\
2003 Feb-Apr & 52730--52747 &$32.8^{+4.1}_{-1.3}$&$0.0^{+20.2}_{-0.0}$&$18.2^{+20.4}_{-0.7}$&$20.4^{+1.2}_{-0.4}$&$20.4^{+9.3}_{-0.4}$\\
2004 Feb-Jun & 53171--53180 &$7.3^{+0.8}_{-0.7}$&$6.1^{+0.1}_{-3.7}$    &$2.0^{+0.5}_{-0.5}$    &$9.6^{+0.5}_{-0.5}$  &$7.0^{+0.9}_{-0.8}$\\
2005 Mar-May & 53494--53510&$2.8^{+0.1}_{-0.1}$&$2.8^{+0.1}_{-0.2}$    &$1.6^{+2.0}_{-1.2}$   &$5.9^{+0.1}_{-0.1}$   &$6.1^{+3.0}_{-3.1}$\\
2005 Nov         & 53695--53704&$3.0^{+0.5}_{-0.5}$&$0.0^{+3.5}_{-0.0}$     &$1.8^{+0.5}_{-1.6}$   &$6.2^{+0.5}_{-0.6}$   &$6.3^{+0.8}_{-4.4}$\\
2006 Jul-Aug  & too few points  & 	--                             &      --                             &       --                          &      --                            &     -- \\
2007 May-Jun & too few points	 & 	--                             &      --                             &       --                          &      --                            &     -- \\
2007 Sep         & 54370--54382&$14.4^{+0.3}_{-0.2}$&$13.8^{+0.4}_{-0.4}$&$1.8^{+0.4}_{-0.4}$   &$13.5^{+0.1}_{-0.1}$&$6.5^{+0.6}_{-0.7}$\\
2008 May-Jun & 54676--54690&$2.0^{+0.1}_{-0.1}$&$2.0^{+0.1}_{-0.2}$    &$5.4^{+2.0}_{-1.4}$   &$5.0^{+0.1}_{-0.1}$   &$11.1^{+1.9}_{-1.5}$\\
2009 Mar-Apr & too few points & 	--                             &      --                             &       --                          &      --                            &     -- \\
2009 Nov-Dec & 55168--55177&$2.8^{+0.2}_{-0.2}$&$2.0^{+0.4}_{-0.4}$    &$2.2^{+0.2}_{-0.2}$    &$6.0^{+0.2}_{-0.3}$  &$7.0^{+0.4}_{-0.4}$\\
2010 Sep-Oct & 55472--55490 &$13.4^{+0.8}_{-0.8}$&$9.1^{+1.2}_{-9.1}$  &$12.0^{+13.7}_{-2.7}$&$13.0^{+0.4}_{-0.4}$&$16.6^{+7.7}_{-2.0}$\\
\hline
\end{tabular}
\end{center}
\end{table*}

\begin{figure}
\begin{center}
\includegraphics[width=6cm,angle=-90]{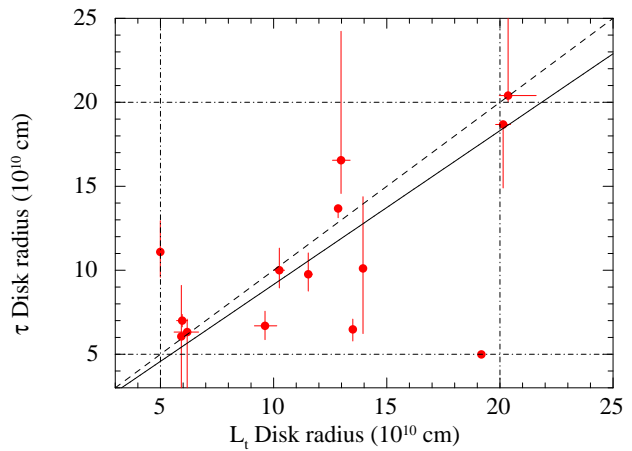}
\end{center}
\caption{Outer disc radius estimate based on the knee luminosity at the exponential/linear transition ($x$ axis) and on the exponential 
decay time ($y$ axis). Horizontal and vertical dot-dashed lines mark the circularisation radius (smaller) and  the distance 
between the neutron star and the inner Lagrangian point (larger). The dashed line indicates the equality among the two radii, whereas the
continuous line is the best fit line.}
\label{enetqui}
\vskip -0.1truecm
\end{figure}

Adopting the values usually reported in the literature for $\nu$ and $\Phi$,  the outer disc radius is smaller than 
circularisation radius.
Imposing the condition that $R_{\rm disc}(\tau_e)$ is larger than the circularisation radius, we can infer that 
$\nu\gsim 9\times 10^{15}$ cm$^2$ s$^{-1}$ for all radii derived from the exponential decay. 
Using this value for $\nu$ all the derived radii are also smaller than the $b_1$ radius, consistent
with expectations (see Fig. \ref{enetqui}).
The same constraint can be applied to the luminosity radius $R_{\rm disc}(L_t)$. In this case we derive 
$\Phi\lsim 8\times 10^{-16}$ cm$^2$ s erg$^{-1}$.
Again, imposing this value, all the radii are also consistent to be smaller than the $b_1$ radius (see Fig. \ref{enetqui}).

Assuming these values for $\nu$ and $\Phi$ one can compare the two determinations of the radii with the two methods 
finding a relatively good agreement, with only a few outliers. Overall there is a factor of $\sim 4$ variation in the estimated 
outer disc radii among different outbursts. This can be due to (small) variations in the outer disc radius, outer disc viscosity, or 
irradiation parameter. 

\section{Discussion and conclusions}

\begin{figure}
\begin{center}
\includegraphics[width=6cm,angle=-90]{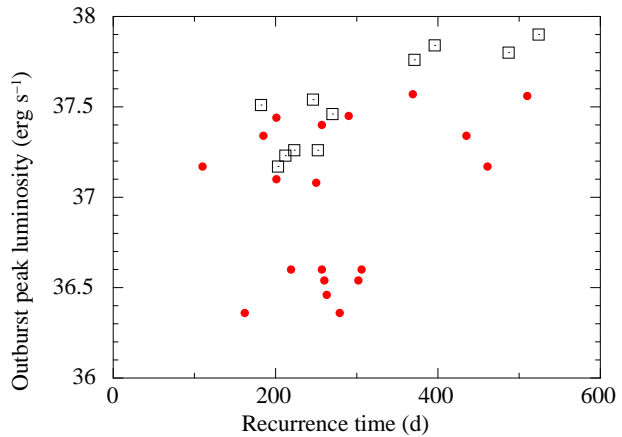}
\end{center}
\caption{Outburst peak X--ray luminosity vs. recurrence time. Red filled circles refer to $RXTE/ASM$ and $MAXI$ 
data during the time interval 1996-2011 (this work). 
Black open squares refer to the time interval 1970-1979 (Kitamoto et al 1993).
 }
\label{outbpeak}
\vskip -0.1truecm
\end{figure}

The outbursts from Aql X-1 observed by $RXTE$/ASM span a long time interval (Jan 1996 -- Sep 2011).
Twenty outbursts were recorded. Outbursts occur mainly in the FRED flavour and often with secondary maxima (see Table 1).
Peak luminosities can be divided into two separated groups: 11 outbursts fall within $10^{37}-10^{37.6}$ erg s$^{-1}$
and 9 within $10^{36.3}-10^{36.6}$ erg s$^{-1}$ with no outbursts in between (factor of $\sim 3$ luminosity gap).
Ideally one would expect a correlation between the outburst peak luminosity and the recurrence time but this is not the case
(see Fig. \ref{outbpeak}). In Fig. \ref{outbpeak} we also included outbursts from {\it Vela 5B} and {\it Ariel 5} (Kitamoto et al. 1993). 
Due to the higher limiting fluxes of these instruments, all the old outbursts fall in the bright group.
A weak correlation ($4\%$ probability,  according to the Spearman rank test) is present among these two variables, 
including the $RXTE$ outbursts and the {\it Vela 5B} and {\it Ariel 5}  ones (see Fig. \ref{outbpeak}). 

\begin{figure}
\begin{center}
\includegraphics[width=6cm,angle=-90]{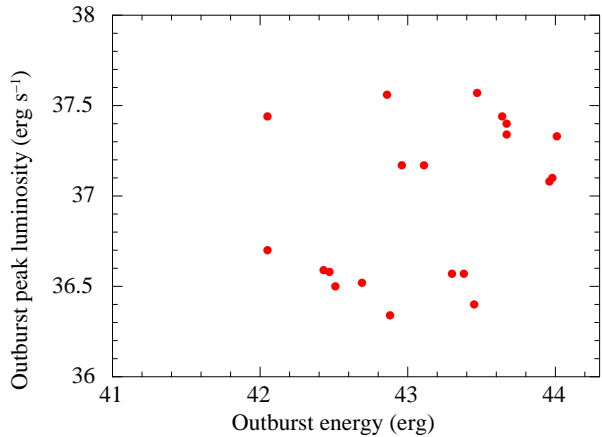}
\end{center}
\caption{Outburst peak X--ray luminosity vs. total outburst energy. Red filled circles refer to $RXTE/ASM$ and $MAXI$ 
data during the time interval 1996-2011. 
 }
\label{lpeakene}
\vskip -0.1truecm
\end{figure}

One can also expect a correlation among the peak outburst luminosity and the total energy of the outburst.
This is not the case as shown in Fig. \ref{lpeakene}. According to a Spearman rank correlation test, there is $23\%$ of a chance 
correlation. This is due to the presence of $\sim 4$ long-lasting low peak luminosity outbursts with a large emitted energy 
as well as $\sim 1$ short-duration high peak luminosity low energetic burst. This indicates that the peak outburst luminosity 
is not a good tracer of the outburst energetics.

We model the outburst decays with a combination of exponential and linear functions following Powell et al.  (2007).
A note of caution should be added since not all the low mass transient outbursts can be described by this model.
A clear example is provided by the 2004 outburst from IGR J00291+5934 showing a bright outburst with a clear 
linear decay (Hartman, Galloway \& Chakrabarty 2011).

Our modeling leads to two different estimates of the outer disc radius depending on the outer disc viscosity and on the outer
disc irradiation properties.  Imposing that the outer disc radius lies between the circularisation radius and the distance 
between the neutron star and the inner Lagrangian point, we can constrain these two parameters. 
In particular, we derive $\nu\gsim 9\times 10^{15}$ cm$^2$ s$^{-1}$ and 
$\Phi\lsim 8\times 10^{-16}$ cm$^2$ s erg$^{-1}$. 
The viscosity estimate is somewhat larger than the value obtained by Powell et al. (2007). These authors found  
$\nu\sim 4\times 10^{14}$ cm$^2$ s$^{-1}$ (factor of $\sim 20$ smaller), considering short orbital period accreting millisecond pulsars. 
The value we derived is also somewhat larger than the one obtained by Shahbaz et al. (1998) for Aql X-1 (factor 3--5),
based, however, on a simpler modeling. 
This might be related to the very long orbital period of Aql X-1, pushing the outer disc edge a factor of 10--100 out with 
respect to the systems considered by Powell et al. (2007).
The only long orbital period system considered by Powell et al. (2007) is the bursting pulsar GRO J1744--28 ($P_{\rm orb}=11.8$ d), 
which is indeed characterised by a factor of $\sim 10$  longer exponential decay time, confirming the increasing trend of the viscosity 
for larger accretion disk outer edges.

The irradiation parameter is instead lower than current estimates: $(4-9)\times10^{-15}$ cm$^2$ s erg$^{-1}$ based on King \& Ritter (1998) 
or $1.3\times10^{-15}$ cm$^2$ s erg$^{-1}$ on Powell et al. (2007). 
Our value is consistent with the latter estimate (just a factor of $\sim 2$ lower), whereas longer orbital period 
systems, such as GRO J1744--28, are characterised by $\Phi\sim 10^{-15}-10^{-14}$ cm$^2$ s$^{-1}$, indicating that 
irradiation effects play minor role.

Adopting our values for $\nu$ and $\Phi$ we found a good agreement among the two different outer radii 
estimates based on the outburst decay light curve fitting. From Fig. \ref{enetqui} it is apparent that, for common values of 
$\nu$ and $\Phi$, the outer radius can span a factor of $\sim 4$ variation among different outbursts.

Given our 20 yr baseline we can compute the mean mass accretion rate of Aql X-1, integrating the observed luminosity.
The mean mass accretion rate is $\mdot\sim 7\times 10^{15}$ g s$^{-1}$ (assuming a standard 
1.4 $\msole$ and 10 km neutron star). Given the 18.97 hr orbital period, this value places Aql X-1 well within 
the unstable region of the Disc Instability Model (Lasota 2001).

Finally, we estimated a mean recurrence time for the system of $280\pm103$ d ($1\,\sigma$). This is considerably longer than the early
recurrence time ($\sim 125$ d) estimated by Priedorsky \& Terrell (1984) and more in line with the estimate of $\sim 300$ d
by Kitamoto et al. (1993; see also \v{S}imon 2002). We also note that dim outbursts (those with a peak luminosity lower than 
$10^{36.6}$  erg s$^{-1}$)  are more regular with a recurrence of $256\pm47$ d ($18\%$ dispersion). With a frequency of 1.36 
outbursts per year in the $RXTE$ era, Aql X-1 is the most prolific neutron star transient in the Galaxy.

\section{Acknowledgments}
We thank the referee for useful comments and suggestions.
We thank Andrea Melandri for comments.

\end{document}